\documentclass[11pt]{article}

\usepackage{graphicx}

\usepackage{amsmath}
\usepackage{amsfonts}
\usepackage{amssymb}

\begin{document}

\pagestyle{empty}

\begin{flushright}
WM-06-102 \\
\end{flushright}

\begin{center}
\vspace*{1.2cm}

{\LARGE\bf Emerging Holography}

\vspace*{1cm}

{\large Joshua Erlich$^a$, Graham D. Kribs$^b$, and Ian Low$^c$}

\vspace*{1cm}
\mbox{$^a$\textit{Department of Physics, College of William and Mary,
                  Williamsburg, VA 23187}} \\[2mm]

\mbox{$^b$\textit{Department of Physics and Institute of Theoretical Science,}}
\\ \mbox{\textit{University of Oregon, Eugene, OR 97403}} \\[2mm]

\mbox{$^c$\textit{School of Natural Sciences, Institute for Advanced Study,
                  Princeton, NJ 08540}} \\[5mm]

\mbox{\texttt{jxerli@wm.edu, kribs@uoregon.edu, ian@ias.edu}}
\vspace*{1cm}

\begin{abstract}
We rederive AdS/CFT predictions for infrared two-point functions
by an entirely four dimensional approach, without reference to
holography. This approach, originally due to Migdal in the context
of QCD, utilizes an extrapolation from the ultraviolet to the
infrared using a Pad\'e approximation of the two-point function.
We show that the Pad\'e approximation and AdS/CFT give the same
leading order predictions, and we discuss including power corrections
such as those due to condensates of gluons and quarks in QCD.  At
finite order the Pad\'e approximation provides a gauge invariant
regularization of a higher dimensional gauge theory in the spirit
of deconstructed extra dimensions.  The radial direction of
anti-de Sitter space emerges naturally in this approach.
\end{abstract}


\end{center}

\newpage
\pagestyle{plain}
\section{Introduction}
\label{sec:introduction}


The AdS/CFT correspondence provides an intriguing relationship
between a weakly coupled $d+1$ dimensional gravity theory and a
strongly coupled $d$ dimensional conformal field theory (CFT)
\cite{Maldacena:1997re,Gubser:1998bc,Witten:1998qj}. The evidence
for a holographic relationship between five-dimensional (5D)
gravity theories and four-dimensional (4D) CFTs is by now
relatively well established, and articulated, for example, in
\cite{Arkani-Hamed:2000ds}. This led to several applications,
particularly with regard to interpreting versions of the
Randall-Sundrum model \cite{Randall:1999ee,Randall:1999vf} as an
approximate 4D CFT whose conformal invariance is broken in the
infrared (IR). The IR breaking of conformal invariance implies a
discrete spectrum of CFT resonances (on the 5D side:  spectrum of
Kaluza-Klein excitations), and thus we have a 4D theory that
really has particles and an $S$-matrix, allowing us to exploit
this relationship for phenomenology.

Recent phenomenological applications of the AdS/CFT correspondence
are the proposals of a simple holographic dual to low energy QCD
\cite{Sakai:2004cn,deTeramond:2005su,Erlich:2005qh,DaRold:2005zs}.
These models have allowed for the calculation of properties of
mesons and baryons, given relatively few input parameters, with
results that so far are remarkably consistent with experimental
data to within 15\%. Physics such as chiral symmetry breaking and
confinement that are associated with the existence of an IR brane
is put in by hand, following the AdS/CFT dictionary. The discrete
spectrum of Kaluza-Klein excitations of bulk fields become the
composites of QCD.  These models are referred to as AdS/QCD
models; several extensions and simplifications of the AdS/QCD
models have been proposed in
\cite{Hirn:2005nr,Sakai:2005yt,DaRold:2005vr,Ghoroku:2005vt,Hirn:2005vk}. 
For
some of the earlier attempts on applying AdS/CFT to QCD, see
\cite{Son:2003et,Brodsky:2003px,Evans:2004ia,Hong:2004sa}. What is
most surprising about the success of AdS/QCD is the success
itself. The AdS/CFT correspondence is a duality between a CFT and
a string theory in a particular spacetime background. If $N_c$ and
$g^2N_c$ are large, where $N_c$ and $g$ are the number of colors
and gauge coupling constant respectively, then stringy physics
decouples and a classical field theory coupled to gravity remains.
However, in QCD neither $N_c$ nor $g^2 N_c$ is large, which
suggests that even if there is a holographic dual to QCD, the full
string theory may be required to make useful predictions.
Nevertheless, it turns out that a simple model using a 5D field
theory in AdS space works surprisingly well.

One of our motivations in this paper is to address why the AdS/CFT
approach has worked so well for QCD, by comparing with an
orthogonal approach that yields similar results. Shifman
\cite{Shifman:2005zn} and Voloshin \cite{voloshin} recently
pointed out that in the '70s Migdal computed masses of mesons in
large $N_c$ QCD using a Pad\'e approximation to QCD
current-current correlators, and found them to be roots of Bessel
functions \cite{Migdal:1977nu,Migdal:1977ut}. (See \cite{Leigh:2005dg}
for spectrum of QCD in 2+1 dimensions involving roots
of Bessel functions.)
The technique
amounts to approximating the current-current correlator in the
deep Euclidean regime by a ratio of two polynomials of degree $N$
(not to be confused with $N_c$ the number of colors). For large
Euclidean momenta, correlators are dominated by short-distance
physics and exhibit conformal behavior. The Pad\'e approximant is
then analytically continued to the low energy regime where
resonance physics dominates.  Physically, the approach is
tantamount to finding the best fit to high energy data with a
finite set of resonances. At low energies and large $N$, Migdal
found that the Pad\'e polynomials could be approximated by Bessel
functions, giving the masses of the mesons as roots of Bessel
functions. The appearance of Bessel functions in this approach is
tantalizingly similar to AdS/CFT.

In this paper we will make explicit the relation between the
Pad\'e approximation of current-current correlators and the
AdS/CFT correspondence, and show that both approaches give
identical leading order predictions. The comparison requires a
careful treatment of the Pad\'e approximation in which a UV
regulator scale $\mu$ and the polynomial degree $N$ are sent to
infinity while the ratio $N/\mu$ is held fixed. The length scale
$N/\mu$ is identified with the location of the IR brane in the
AdS/CFT approach, which is also related to the confinement scale.
We emphasize that just like in AdS/QCD, the IR scale is put in by
hand, and is {\em a priori} independent of the chiral symmetry
breaking condensates. Nevertheless, the Pad\'e approximation is
remarkably powerful in being able to manifest other aspects of
AdS/CFT.  For instance, a very recent paper \cite{Hirn:2005vk}
incorporated power corrections arising from condensates of quarks
and gluons \cite{Shifman:1978bx,Shifman:1978by} in the AdS/QCD
framework. We show that the Pad\'e approximation can also
accommodate power corrections, and the result continues to agree
with the AdS/CFT approach.  Finally, we show how the radial
direction of 5D anti-de Sitter space emerges as the order of the
Pad\'e approximation increases, by analogy with deconstructed
extra dimensions \cite{Arkani-Hamed:2001ca,Hill:2000mu}.

The outline of this paper is as follows.  In Sec.~\ref{sec:AdSCFT}
we review the computation of conformal two-point functions in
AdS/CFT.  In Sec.~\ref{sec:midgal}, we describe Migdal's Pad\'e
approximation for the current-current correlation function and
show that in the low-energy, large $N$ limit it gives the same
expression as that obtained from AdS/CFT.  In Sec.~\ref{sec:power}
we consider the breaking of the conformal symmetry through power
corrections, and again show that the two approaches agree.  In
Sec.~\ref{sec:deconstruct} we show how the Pad\'e approximation
gives rise to a gauge invariant regularization of warped 5D gauge
theories by analogy with deconstructed extra dimensions. Finally,
we conclude in Sec.~\ref{sec:conclusions}.

\section{The AdS/CFT Correspondence}
\label{sec:AdSCFT}

The prototypical example of the AdS/CFT correspondence is the
duality between 4D ${\cal N}$=4 $SU(N_c)$ super Yang-Mills theory
and Type IIB string theory compactified on AdS$_5\times $S$_5$. In
a limit which involves taking the rank of the gauge group and the
't Hooft coupling large, string theory effects decouple and
classical Type IIB supergravity on AdS$_5\times$S$_5$ remains.
With this simplification it becomes possible to analytically
calculate correlation functions in the 4D gauge theory from
supergravity.  The basic dictionary relating correlation functions
in the CFT to supergravity on an AdS background was found in
\cite{Gubser:1998bc,Witten:1998qj}, and many tests have since been
performed. In this section we briefly review the computation of
the CFT two-point function for vector currents in AdS/CFT. Each
operator in the CFT corresponds to a field in anti-de Sitter space
of one dimension higher, whose boundary value is set equal to the
source for the corresponding operator. More precisely, if we
consider the AdS metric in the form,
\begin{equation}
ds^2 = \frac1{z^2}\left(\eta_{\mu\nu} dx^\mu dx^\nu - dz^2\right),
\end{equation}
where $\eta_{\mu\nu} = {\rm Diag}(1,-1,-1,-1)$,
the prescription in \cite{Gubser:1998bc,Witten:1998qj} states that
\begin{equation}
\label{pres}
\langle e^{ i \int d^4x\, \phi_0(x){\cal O}(x)} \rangle _{\rm CFT}
   = \left. e^{i S_{\rm AdS}[\phi^{\rm cl}]}\right|_{\phi^{\rm cl}(x,z=0)=
\phi_0},
\end{equation}
where $S_{\rm AdS}[\phi^{\rm cl}]$ is the classical action in the AdS space
and $\phi^{\rm cl}$ is a solution to the equation of motion whose boundary
value is fixed to be the source $\phi_0(x)$ (up to a conformal rescaling).
In practice one introduces an UV regulator
$z=\epsilon$ and then takes $\epsilon\to 0$. For a vector
current such as $J^\mu=\bar{q}\gamma^\mu q$ in the CFT, there is a corresponding
bulk gauge field $A_M(x,z)$ whose boundary value is the source for
$J^\mu$. The 5D action in AdS space is
\begin{equation}
S_{\rm AdS} = -\int d^4x dz \sqrt{-g}\ \frac{1}{4g_5^2} F_{MN} F^{MN} ,
\end{equation}
where the capital roman letters
$M,N = 0,1,2,3,z$. We have neglected a bulk mass term for $A_M$ since we
are considering conserved currents.
According to the AdS/CFT correspondence, to calculate the current-current
correlator, we calculate the 5D action on a solution to the equations of
motion for the corresponding gauge field such that the 5D gauge field at
the UV boundary has the profile of the 4D source of the current. We will
consider a finite AdS space with an infrared boundary at $z=z_0$.
The profile of the 5D gauge field satisfying those boundary conditions
is the bulk-to-boundary propagator, which we call $V(q,z)$.
Varying the action (twice) with respect to the boundary source gives the
current-current correlator.  We impose the $A_z=0$ gauge and
Fourier-transform the gauge field in four dimensions:
\begin{equation}
\label{bb0}
A_\mu(q,z)=\frac1{V(q,\epsilon)}\widetilde{A}_\mu(q)V(q,z),
\end{equation}
where $\widetilde{A}_\mu(q)$ is the Fourier-transformed current
source. The boundary condition
$A_\mu(q,\epsilon)=\widetilde{A}_\mu(q)$ is built into
(\ref{bb0}). The boundary condition at $z=z_0$ is not completely
predetermined, but for definiteness we assume Neumann boundary
conditions there, $\partial_z V(q,z_0)=0$, corresponding to the
gauge invariant condition $F_{\mu z}(x,z_0)=0$. The equations of
motion for the transverse part of the gauge field are,
\begin{equation}
z\,\partial_z \left(\frac1z \partial_z V(q,z)\right)+q^2\,V(q,z) = 0,
\end{equation}
which, given the boundary conditions, leads to
\begin{equation}
\label{btob} V(q,z) =q z \left(Y_0 (q z_0) J_1 (q z) - J_0(q z_0)
Y_1(q z)\right).
\end{equation}
Evaluating the action on the solution leaves only the the boundary term
at the UV
\begin{equation}
\label{act0}
S_{\rm AdS} = -\frac1{2g_5^2} \int d^4q\, \widetilde{A}_\mu(q) \widetilde{A}^\mu(-q)
  \left(\frac1{z} \frac{\partial_z V(q,z)}{V(q,z)} \right)_{z=\epsilon}.
\end{equation}
If we write the Fourier-transformed vector current two-point function as,
\begin{equation}
\int d^4x\, e^{i q\cdot x}\, \langle J_\mu(x) J_\nu(0) \rangle
= \left( g_{\mu\nu} - \frac{q_\mu q_\nu}{q^2} \right) \Sigma(q^2),
\end{equation}
then functionally differentiating the action (\ref{act0}) with respect to the source
$\widetilde{A}_\mu(q)$  yields the vector current-current correlator
determined by AdS/CFT:
\begin{eqnarray}
\label{2pt} \Sigma(q^2)&=& - \frac{1}{g_5^2}\left.\frac1{z}\
  \frac{\partial_z V(q,z)}{V(q,z)}\right|_{z=\epsilon
\rightarrow 0} \nonumber \\
                   &=& -\frac{1}{g_5^2}
\left. \frac1z\ q\  \frac{Y_0 (q z_0) J_{0} (q z) - J_0(q z_0) Y_{0}(q z)}
                                       {Y_0(q z_0) J_1(q z) - J_0(q z_0) Y_1(q z)}\right|_{z=\epsilon} \nonumber \\
                   &\to& \frac{q^2}{g_5^2}
                    \frac{ \log (q\epsilon)J_0(q z_0) - (\pi/2)Y_0(q z_0)}{J_0(q z_0)},
\end{eqnarray}
where we have retained the leading non-vanishing contribution from each of the Bessel functions in
the limit $q \epsilon \to 0$:
\begin{equation}
J_0(q\epsilon) \to 1, \quad J_1(q\epsilon) \to \frac{q\epsilon}2
\to 0, \quad Y_0(q\epsilon) \to \frac2{\pi} \log (q\epsilon) ,
\quad Y_1(q\epsilon) \to - \frac2{\pi} \frac1{q\epsilon}.
\end{equation}
Note that we have neglected terms in (\ref{2pt}) that can be
absorbed into the re-definition of $\epsilon$, such as the Euler
constant $\gamma_{E}$ in the expansion of $Y_0(q\epsilon)$.
 From (\ref{2pt}) we see that the masses of the
resonances in the CFT, given by the simple poles in $\Sigma(q^2)$,
are determined by roots of the Bessel function $J_0(qz_0)$. On the
other hand, if we simultaneously take $qz_0\gg 1$, then the
spectrum becomes continuous and the vector polarization
$\Sigma(q^2)$ turns into, up to contact terms,
\begin{equation}
\label{sig0}
\Sigma(q^2)\rightarrow \frac{1}{2g_5^2}\,q^2\log(q^2\epsilon^2).
\end{equation}
Here we see explicitly that $\epsilon$ plays the role of an UV regulator.
In the context of AdS/QCD, one matches (\ref{sig0}) onto
 the quark bubble calculation in QCD at large Euclidean momentum, which gives,
\begin{equation}
\label{g5}
g_5^2=\frac{12\pi^2}{N_c}
\end{equation}
in units of the AdS curvature \cite{Erlich:2005qh,DaRold:2005zs}.

The expression in (\ref{g5}) serves as a good point to discuss why a
strict reading of the AdS/CFT correspondence would suggest that
the models in AdS/QCD shouldn't have worked so well. In QCD
$N_c=3$ which then suggests that the 5D gauge coupling is strong.
Therefore higher order corrections in the AdS side may be as
important as the tree-level results.  Furthermore, on the QCD side the
't Hooft coupling $g_4^2N_c \sim 1$, which would suggest stringy
effects cannot be decoupled in the AdS space. Again, a simple model
of a 5D gauge field does not appear to be justified {\em a priori}, yet
somehow produces accurate results.

In order to include corrections due to higher dimension operators
in the operator product expansion (OPE) of the two-point function
(power corrections), we can add to the action higher dimension
operators and additional bulk fields
\cite{Erlich:2005qh,DaRold:2005zs}.  For example, including the
dilaton in the theory allows couplings which would mimic the
inclusion of higher dimension operators involving the QCD field
strength.  Alternatively, a simpler modification of the theory
which also captures the effects of power corrections in the
two-point function is to allow deviations in the geometry towards
the boundary \cite{Hirn:2005vk}.  These modifications directly
encode breaking of conformal symmetry away from the UV, which is
also a consequence of chiral symmetry breaking.  We will see in
Sec.~\ref{sec:power} that Migdal's Pad\'e approximation approach
to estimating current correlators can be easily generalized to
include higher dimension operators in the OPE.

\section{The Pad\'e Approximation and Vector Mesons}
\label{sec:midgal}

In the '70s Migdal proposed a systematic procedure for calculating
masses of mesons in large $N_c$ QCD
\cite{Migdal:1977nu,Migdal:1977ut}, which approximates the
two-point function in the deep Euclidean region by a rational
function and then analytically continues into the infrared domain.
Mathematically this procedure amounts to performing the Pad\'e
approximation. The rationale behind  Migdal's approach relies on
properties of large $N_c$, in particular the vanishing of
instanton corrections to current correlators.  It is unclear how
nonperturbative corrections should be added in the 1/$N_c$
expansion, so the justification for the Pad\'e approximation is
limited. Nonetheless, ignoring the effects of QCD condensates,
Migdal found that the masses of the mesons in large $N_c$ QCD are
proportional to roots of certain Bessel functions, which is quite
intriguing from the modern perspective, as was recently emphasized
by Shifman \cite{Shifman:2005zn} and Voloshin \cite{voloshin}.

It is the purpose of this section to first give an overview of
Migdal's approach from nearly three decades ago and make precise
the relation between AdS/CFT and the Pad\'e approximation of
correlation functions. We will focus our discussion on the
two-point function of conserved vector currents in a CFT, although
the construction can be generalized to arbitrary conformal tensors
\cite{Migdal:1977nu,Migdal:1977ut}.
Two-point correlators of operators with arbitrary conformal dimension in the
UV generally involve expressions of the form,
\begin{equation}
\label{2pt1}
\Sigma_\nu(q^2) = (q^2)^\nu \log \frac{q^2}{\mu^2}.
\end{equation}
The scale $\mu$ is a renormalization scale and $\nu=\Delta-2$, where $\Delta$ is the
conformal dimension of the operator involved. For a conserved vector current,
 $\Delta=3$
and $\nu=1$, which is consistent with (\ref{sig0}).
We will consider the dimensionless quantity
\begin{equation}
\label{fnu}
f_0(t) = \log t
\end{equation}
so that $f_0(q^2/\mu^2) = (q^2)^{-\nu} \Sigma_\nu(q^2)$.  Migdal
proposed that one find a ratio of two polynomials of degree $N$
respectively, which reproduces the first $2N+1$ terms in the
Taylor expansion of  $f_0(t)$ with respect to an arbitrary
subtraction point, which we choose to be $t=-1$, {\em i.e.}
$q^2=-\mu^2$.
Mathematically this procedure is known as the Pad\'e
approximation, which allows functions with possible poles and
branch cuts to be well-represented by polynomials.

\begin{figure}
\centering
\includegraphics[scale=0.5]{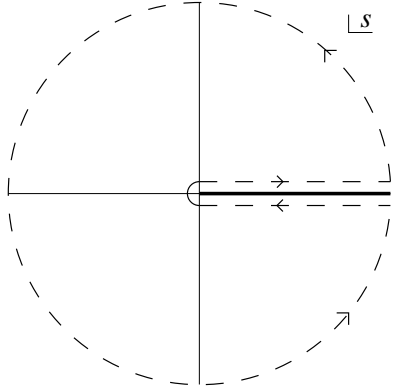}
\caption{\label{fig:Cauchy}Contour used to evaluate derivatives of
$Q_N f_0(t)$ in (\ref{eq:Cauchy}).}
\end{figure}

To be precise, we are looking for two polynomials $R_N(t)$ and $Q_N(t)$ of degree $N$
 that satisfy
\begin{equation}
\label{pade0}
f_0(t)  - \frac{R_N(t)}{Q_N(t)} = {\cal O}\left((t+1)^{2N+1}\right),
\end{equation}
which in turn implies
\begin{equation}
\label{pade} \left. \frac{d^m}{dt^m}  \left[ Q_N f_0(t)  - R_N
\right] \right|_{t=-1} = 0, \quad
   m=0,1,\cdots,2N.
\end{equation}
Notice that, since $R_N(t)$ is only of degree $N$, it is
determined completely by the first $N+1$
equations in (\ref{pade}) for $m=0,1,\cdots, N$, if $Q_N$ is known;
the remaining $N$ equations determine
the denominator function $Q_N(t)$ up to an overall constant (corresponding
to a simultaneous rescaling of $R_N$ and $Q_N$):
\begin{equation}
\label{qn} \left. \frac{d^m}{dt^m} Q_N f_0(t) \right|_{t=-1} = 0,
\quad m = N+1, \cdots, 2N.
\end{equation}
To solve for $Q_N$
we can analytically continue (\ref{qn}) into the complex plane.
The function $f_0(t)$ has a branch cut which is taken to be along the positive
real axis.  This branch cut is approximated by a series of poles in the
Pad\'e approximation, and the resonance masses are interpreted as the
locations of the poles.  The imaginary part of the function $f_0(t)$ jumps
by $2\pi$ across the branch cut.  At any location in the complex $t$-plane
away from the positive
real axis, derivatives of the function $Q_N f_0(t)$
are given by the Cauchy integral formula along the path in
Fig~\ref{fig:Cauchy}.
The integral over the large and small circles vanish (for the
orders of derivative we are considering), and the integrals above
and below the branch cut cancel except for the change in the
imaginary part of $f_0$.  What remains is given by,
\begin{equation}\label{eq:Cauchy}
\left. \frac{d^m}{dt^m} Q_N f_0(t) \right|_{t=-1}
  = \frac{m!}{2\pi} \int_0^{\infty} ds \ \frac{Q_N\, \Delta{\rm Im} f_0(s)}{
(s+1)^{m+1}} = 0,\quad
  m=N+1, \cdots, 2N,
\end{equation}
where $\Delta{\rm Im} f_0(t)={\rm Im} f_0(t+i\epsilon)-{\rm Im}f_0(t-i\epsilon)$ is
the change in $f_0(t)$ across the branch cut.
On the
other hand, from (\ref{pade}) we have
\begin{eqnarray}
\label{pm} R_N &=& Q_N f_0(t) - \sum_{n=N+1}^{\infty} \frac1{n!}
\left.\frac{d^n}{dt^n} Q_N f_0(t) \right|_{t=-1}\,
(t+1)^n \nonumber \\
   &=& Q_N f_0(t) - \sum_{n=N+1}^{\infty} \int_0^{\infty} ds\ \frac{1}{2\pi}
                             \frac{Q_N\,\Delta{\rm Im} f_0(s)}{(s+1)^{n+1}}
                 (t+1)^n \nonumber \\
     &=& Q_N f_0(t) - \int_0^{\infty} ds\ \frac1{2\pi} \frac{Q_N\,\Delta
                   {\rm Im}f_0(s)}{s-t}
                    \left(\frac{t+1}{s+1}\right)^{N+1}.
\end{eqnarray}
The second term on the right hand side cancels the higher order
terms in the Taylor expansion of $Q_N f_0(t)$ so that the result
is a polynomial of degree $N$. In our case, $\Delta {\rm Im}f_0=-2\pi$, 
and from (\ref{qn}) and (\ref{eq:Cauchy}) we obtain
the following equations for $Q_N$:
\begin{equation}
\int_0^{\infty} ds \frac{Q_N}{(s+1)^{m+1}} = 0, \qquad m=N+1,
\cdots, 2N \; .
\end{equation}
This set of conditions for $Q_N$ is most transparent after doing
the change of variables $s = (1 - x)/(1 + x)$, 
\begin{equation}
\int_{-1}^{1} d x \, (1 + x)^{m - 1} Q_N\left( \frac{1-x}{1+x} \right) = 0 \; ,
\end{equation}
and shifting $m \to m^\prime=m - N - 1,$
\begin{equation}
\label{qn1}
\int_{-1}^{1} d x \, (1+x)^{m'} (1+x)^N Q_N\left(\frac{1-x}{1+x} \right) = 0 
\; , \qquad m'=0, 1, \cdots, N-1.
\end{equation}
This equation contains a product of two polynomials.  The first one
can be written as a sum over Legendre polynomials
\begin{equation}
(1 + x)^{m'} = \sum_{i=0}^{m'} c_i P_i(x) \; ,
\end{equation}
where $c_i$ are calculable coefficients (that we do not need).  
The second piece 
\begin{equation}
(1 + x)^N Q_N\left( \frac{1-x}{1+x} \right) 
\label{second-piece}
\end{equation}
is some polynomial of degree $N$ in $x$.  This can be seen by realizing 
that $Q_N\left((1-x)/(1+x)\right)$ is a polynomial of degree $N$ in $(1-x)/(1+x)$, 
and so the factor $(1+x)^N$ removes all powers of $(1+x)$ in the denominator 
of the expansion of $Q_N$.  Legendre polynomials satisfy the orthogonality 
relation
\begin{equation}
\int_{-1}^{1} d x \, P_a(x) P_b(x) = 0 \quad \mbox{for} \; a \neq b
\end{equation}
which combined with (\ref{qn1}) implies that we can identify the
second piece (\ref{second-piece}) as
\begin{equation}
(1 + x)^N Q_N\left( \frac{1-x}{1+x} \right) = P_N(x) \; .
\label{ident1-eq}
\end{equation}
This identification is unique since (\ref{qn1}) with $m' = N - 1$ 
implies (\ref{second-piece}) must be a polynomial of at least degree $N$,
but we already showed that (\ref{second-piece}) is a polynomial of degree 
exactly $N$.  Rewriting (\ref{ident1-eq}) in terms of the original 
variable $s$ we obtain
\begin{equation}
\label{qn2}
Q_N(s) = (s+1)^N P_N\left(\frac{1-s}{1+s}\right) \; ,
\end{equation}
up to an overall normalization.  The normalization does not affect 
the computation of the two-point function since it is canceled
between the numerator and the denominator in (\ref{pade0}).

Next, in order to make the Pad\'e approximation in (\ref{pade0}) 
as accurate as possible, Migdal proposed taking $N\to \infty$ and 
also the low energy limit $q^2 \ll \mu^2$, in which case the 
Legendre polynomial reduces to a Bessel function.  This is most
easily seen by rewriting the Legendre polynomial as a special
case of a Jacobi polynomial $P_N(x) = P_N^{(0,0)}(x)$ and using 
the relation \cite{szego},
\begin{equation}
\label{bessel}
\lim_{N\to \infty} N^{-a} P_N^{(a,b)}\left(\cos \frac{z}{N}\right) =
     \lim_{N\to\infty} N^{-a} P_N^{(a,b)}\left(1-\frac{z^2}{2N^2}\right)
         = \left(\frac{z}2\right)^{-a} J_a(z).
\end{equation}
Therefore, in this limit with $(a,b) \rightarrow (0,0)$ we have
\begin{equation}
\label{qn3}
Q_N(t) \to J_0(2N\sqrt{t}) \equiv Q_\infty(t).
\end{equation}
For completeness we have provided a derivation of (\ref{qn3}) in
the appendix. Here we see that, if we think of $\mu$ as the UV
cutoff of the theory, then in the low energy and $N\to \infty$
limit, a new, infrared scale emerges as $\mu/N$. This infrared
scale $\mu/N$ is not determined by the Pad\'e approximation.
Instead, it acts as a sort of boundary condition: we can take both
the UV cutoff $\mu$ and the degree $N$ to infinity while keeping
their ratio $\mu/N$ fixed,
\begin{equation}
\label{limits-eq} \mu\to \infty, \quad N\to \infty,  \quad
\frac{\mu}{N} = \mu_{ir}= {\rm fixed} \ .
\end{equation}
In Sec.~\ref{sec:deconstruct} we will give these scales a concrete
physical interpretation in the deconstructed picture.

We can also calculate the numerator in the Pad\'e approximation using (\ref{pm})
\begin{eqnarray}
\label{pn}
R_\infty(t) &\equiv& Q_\infty(t) f_0(t) - \frac1{2\pi}\int_0^{\infty} ds\ 
\frac{Q_\infty(s)\,\Delta
{\rm Im}f_0(s)}{s-t}
              \nonumber \\
   &=&  Q_\infty(t) f_0(t) - \frac1{2\pi}\int_0^{\infty} ds\ \frac{(-2\pi)  I_0(2N\sqrt{-s})}{s-t}
      \nonumber \\
  &=&  J_0(2N\sqrt{t}) (\log t) -  \pi\, Y_0(2N\sqrt{t}).
\end{eqnarray}
Up to an overall constant which has been factored out of (\ref{2pt1}),
the Pad\'e approximation for the polarization $\Sigma(q^2)$ becomes,
\begin{equation}
\label{2pt_mig}
\Sigma(q^2) \propto q^2 \frac{ J_0(2Nq/\mu) \log (q^2/\mu^2)
                   - \pi\, Y_0(2Nq/\mu)}
                { J_0(2Nq/\mu)}.
\end{equation}
Here we see that if we  identify
\begin{equation}
\label{pade1}
\frac1{\mu} = \epsilon , \quad \frac2{\mu_{ir}} = z_0,
\end{equation}
we recover exactly (\ref{2pt}). What is interesting here is that
we have obtained the same results as AdS/CFT for the two-point
function, without reference to the anti-de Sitter space and
holography. Furthermore, in the bulk-to-boundary propagator the
correct dependence on the infrared boundary is reproduced in the
Pad\'e approximation in the limit $N\to\infty$ and
$q^2/\mu^2\ll1$. We should emphasize that the $N\to \infty$ limit
is different from the large $N_c$ limit in a gauge theory; in
fact, Migdal's starting point is already QCD in the large $N_c$
limit.

It is instructive to look at the bulk-to-boundary propagator (\ref{btob}) in AdS/CFT
more closely, paying attention to the analytic
structure of the Bessel functions. The Bessel function of the first kind, $J_\nu(x)$, is
analytic and has no poles in the complex plane. Near the origin it behaves like
$J_\nu(x) \sim x^\nu$. On the other hand, the Bessel function of the second kind
$Y_\nu(x)$ has the following series expansion, when $\nu$ is an integer,
\begin{equation}
Y_\nu(x) \sim \sum_{m=0}^{\nu-1} a_m \frac{x^{2m}}{x^\nu} +
\frac2{\pi} J_\nu(x) \log \frac{x}2
   +\sum_{m=0}^\infty b_m x^{2m+\nu} \quad.
\end{equation}
That is $Y_\nu(x)$ has a pole of order $\nu$ at the origin and a
non-analytic piece $J_\nu(x) \log x$. Thus we see in (\ref{btob})
the terms that are non-analytic in momentum $q$ conspire to cancel
each other. Both the numerator and denominator of the two-point
function determined by AdS/CFT as in (\ref{2pt}) are analytic in
the momentum $q$. AdS/CFT is secretly constructing the Pad\'e
approximation to the two-point function. The reason for this goes
back to the large $N_c$ limit for which quantum corrections can be
ignored in AdS/CFT.  There is no source of non-analyticity at
large $N_c$, and resonances become simple poles in the two-point
function.  What we have found is that AdS/CFT gives the best
approximation to the conformal behavior in the UV by means of a
set of infinitely narrow resonances.

\section{Power Corrections to the Conformal Correlator}
\label{sec:power}

In this section we show that, after breaking the conformal
symmetry by adding power corrections to the two-point function in
(\ref{2pt1}), Pad\'e approximation continues to reproduce the
results of AdS/CFT. The motivation for such considerations stems
from the well-known fact that in the deep Euclidean region one can
perform OPE to the two-point function of vector currents in QCD,
which exhibits a conformal behavior at leading order in $q^2$.
Moreover, at subleading order the OPE contains power corrections
arising from various quark and gluon condensates
\cite{Shifman:1978bx,Shifman:1978by}. In the original publications
\cite{Migdal:1977nu,Migdal:1977ut} Migdal only considered breaking
of the conformal symmetry due to the running of $\alpha_s(\mu)$,
the QCD coupling constant. We will see how to incorporate these
power corrections in both the AdS/CFT and the Pad\'e
approximation. In the context of AdS/QCD, such power corrections
were recently considered in \cite{Hirn:2005vk} from a geometric
perspective, which we will follow\footnote{We are grateful to
Veronica Sanz for kindly clarifying the computation in
\cite{Hirn:2005vk}.}.

Power corrections arise from condensates of operators, so AdS/CFT
instructs us to turn on a normalizable background for the 5D
fields which act as the source for those operators
\cite{Balasubramanian:1998sn,Balasubramanian:1998de}.  In order to
reproduce the power corrections to correlators in the UV, we can
add appropriate higher dimension operators to the 5D Lagrangian,
and then determine the consequences of these corrections in the IR
\cite{Erlich:2005qh,DaRold:2005zs}.  At lowest order in the chiral
condensates, in \cite{Hirn:2005vk} it is shown that power
corrections can be included in AdS/CFT instead by power law
deviations from the AdS metric near the UV boundary. More
specifically, we consider the metric,
\begin{eqnarray}
ds^2 &=& w(z)^2 \left( \eta_{\mu\nu} dx^\mu dx^\nu - dz^2 \right) \\
w(z) &=& \frac1{z} \left( 1 + O_4^\prime\, z^4 + O_6^\prime\, z^6 + \cdots \right) ,
\end{eqnarray}
where $O_{2n}^\prime$ is a quantity with mass dimension $2n$ and is
assumed to be small
compared to the AdS length scale so that one can perform a
perturbative expansion in $O_{2n}^\prime$.
The bulk-to-boundary propagator $V(q,z)$
satisfies the equation,
\begin{equation}
\left(-q^2 - \partial_z^2 -\frac{w'(z)}{w(z)} \partial_z \right) V(q,z)=0
\end{equation}
with the IR boundary condition chosen to be $\partial_z V(q,z_0)=0$ as before,
while the UV
boundary condition fixes $V(q,\epsilon)$ to some nonvanishing value. (Recall
that the overall normalization of $V(q,z)$ factors out of $\Sigma(q^2)$.)
For $w(z)=w_0(z)=1/z$,
the solution $V_0(q,z)$ is given by (\ref{btob}). We are interested
in the following metric
\begin{eqnarray}
w(z)&=&\frac1{z} \left(1+ O^\prime_{2n}\, z^{2n}\right)\nonumber \\
    &=& w_0(z) \left(1+ w_{2n}(z) \right).
\end{eqnarray}
Writing $V(q,z)=V_0(q,z)+V_{2n}(q,z)$, a perturbative expansion in
$O^\prime_{2n}$ leads to
\begin{equation}
\left(-q^2 - \partial_z^2 - \frac{w_0^\prime(z)}{w_0(z)} \partial_z
  \right) V_{2n} = - w_{2n}^\prime(z) \partial_z V_0.
\end{equation}
Making the change of variable $z \to x = q z$, $V_0(q,z)=V_0(x)$, and
$V_{2n}(q,z) \to \tilde{V}_{2n}(x) = (q^{2n}/O_{2n}^\prime) V_{2n}(q,z)$, we have
\begin{equation}
\label{pert}
\left(1 + \partial_x^2 - \frac{1}{x}\partial_x \right) \tilde{V}_{2n} =
  2n\, x^{2n-1}\, \partial_x V_0.
\end{equation}
A particular solution to the above can be obtained by first solving for
the Green's function $G(x,x^\prime)$
\begin{equation}
\left(1 + \partial_x^2 - \frac{1}{x}\partial_x \right) G(x,x^\prime)
  = x\, \delta(x-x^\prime),
\end{equation}
subject to the boundary conditions: $G(q \epsilon,x^\prime) =
\partial_{x^\prime} G(x,q z_0)=0$. The UV boundary condition is chosen so that
$V(q,\epsilon)$ is unchanged by the perturbation.  We can
solve $G(x,x^\prime)$ in the regions $x<x^\prime$ and $x>x^\prime$
separately, and then match over the delta function, which results in
\cite{Randall:2001gb}
\begin{equation}
\label{prop0}
G(x,x^\prime) = \frac{\pi}2  \frac{x x^\prime}{AD-BC}
\left[ A\, J_1(x_<)+B\, Y_1(x_<) \right] \left[ C\, J_1(x_>) + D\, Y_1(x_>) \right]
\end{equation}
with $x_{<,>}=\{{\rm min,max}\}(x,x^\prime)$ and
\begin{equation}
A=Y_1(q\epsilon),\quad B=-J_1(q\epsilon),\quad C=Y_0(q z_0),\quad D=-J_0(q z_0).
\end{equation}
A particular solution to (\ref{pert}) is then
\begin{eqnarray}
\label{tildv}
\tilde{V}_{2n}(x)&=&\int_0^{x_0} dx^\prime\ 2n (x^\prime)^{2n-2}
(\partial_{x^\prime}V_0)\ G(q z,x^\prime) \nonumber \\
 &=&  \frac{\pi}2 \frac{V_0(x)}{AD-BC}
   \int_0^x dx^\prime\ 2n (x^\prime)^{2n-2}
 (\partial_{x^\prime}V_0)\ x^\prime [ A J_1(x^\prime) + B Y_1(x^\prime)]
   \nonumber \\
   && \quad + \frac{\pi}2 \frac{x [A J_1(x)+B Y_1(x)]}{AD-BC}
        \int_x^{x_0}  dx^\prime\ 2n (x^\prime)^{2n-2}
(\partial_{x^\prime}V_0)\ V_0(x^\prime),
\end{eqnarray}
where $x_0\equiv qz_0$.
Recall that in computing the two-point function (\ref{2pt}) we are only
interested in the behavior of $V(q,z)$ near the UV boundary, so we focus on the limit
$q\epsilon \to 0$ and $x\ll 1$. In this limit,
\begin{equation}
AD-BC \to
-\frac2{\pi} \frac1{q\epsilon} D=
\frac2{\pi} \frac{J_0(qz_0)}{q\epsilon}
        , \quad x [A J_1(x) +B Y_1(x)] \to
         -\frac2{\pi} \frac1{q\epsilon} x J_1(x) \approx -\frac{x^2}{\pi} \frac1{q\epsilon}  .
\end{equation}
Moreover, if we further consider the energy regime where $x_0 \gg 1$,
the leading contribution in (\ref{tildv}) is
\begin{eqnarray}
\label{tildv1}
 \tilde{V}_{2n}(x) &\approx& - \frac{\pi}4 \frac{x^2}{J_0(qz_0)}
\int_0^\infty dx^\prime \ 2n (x^\prime)^{2n-2}
        (\partial_{x^\prime} V_0) V_0(x^\prime)  \nonumber \\
        &=&  -\frac{\alpha}{2\pi} \ x^2 J_0(q z_0),
\end{eqnarray}
where
\begin{eqnarray}
\alpha &=& \frac{\pi^2}{2J_0(qz_0)^2} \int_0^\infty  dx^\prime \ 2n (x^\prime)^{2n-2}
        (\partial_{x^\prime} V_0) V_0(x^\prime)  \nonumber \\
        &=& -\frac{\pi^2}{2J_0(qz_0)^2}\ \frac12
               \int_0^\infty  dx^\prime \ 2n(2n-2)  (x^\prime)^{2n-3}
                 V_0(x^\prime)^2  \nonumber \\
         &=&   (-1)^{n}\ \frac{\sqrt{\pi}\ n^2 \Gamma(n)^3}
                         {\Gamma\left(\frac{1}2+n\right)} .
\end{eqnarray}
Apart from the factor of  $(-1)^n$, $\alpha$ results in the same numerical coefficient
as
in \cite{Hirn:2005vk}, which, instead of solving for the
propagator (\ref{prop0}),  used the ansatz $\tilde{V}_{2n}
= C(x) V_0(x)$ in (\ref{pert}) and then solved for $C(x)$ \cite{sanz}.
The extra
factor $(-1)^n$ arises from the fact that we are considering Minkowski momentum,
whereas \cite{Hirn:2005vk} considered the Euclidean momentum. It is worth noting
that $\tilde{V}_{2n}$, and hence $V_{2n}(q,\epsilon)$,
has the same infrared coefficient, $J_0(qz_0)$,
as $V_0(q,\epsilon)$, which implies that the position of the poles
in the two-point function $\Sigma(q^2)$
is not altered by the power corrections. That is the mass of the
resonances in the CFT is not sensitive to the power corrections
at this order in perturbation theory and in the limit $x_0=qz_0 \gg 1$.
Factoring out a $1/g_5^2$, the two-point function now becomes
\begin{eqnarray}
\label{2pt4}
\Sigma(q^2)
 &\approx& w_0(1+w_{2n}) \left. \left\{\frac{\partial_z V_0(q,z)}{V_0(q,z)}+
\frac{\partial_z V_{2n}(q,z)}{V_0(q,z)} - \frac{\partial_z V_0(q,z)}{V_0(q,z)}\,
\frac{V_{2n}(q,z)}{V_0(q,z)}
   \right\}
  \right|_{z=\epsilon} \nonumber \\
  &\to& \frac1{\epsilon} \left\{\frac{\partial_z V_0(q,\epsilon)+
\partial_z V_{2n}(q,\epsilon)}
   {V_0(q,\epsilon)}
       \right\}   \nonumber \\
 &=& \frac{q^2}2 \frac{\log (q^2\epsilon^2) J_0(qz_0) - \pi Y_0(qz_0)
       -(1/q^{2n}) \alpha\, O^\prime_{2n}\, J_0(qz_0)}{J_0(qz_0)}
 \end{eqnarray}
where the power correction $1/q^{2n}$ comes from only
the $w_0 (\partial_z V_{2n})/V_0$ term. All the other terms involving $w_{2n}$ or $V_{2n}$
vanish in the limit $q\epsilon \to 0$.

Next we discuss how to incorporate the effect of power corrections
in Migdal's regularization, which was not discussed in
\cite{Migdal:1977nu,Migdal:1977ut}. As a matter of fact, it
requires almost no effort to construct the Pad\'e approximation
when including the power corrections to the conformal correlator
(\ref{2pt1}), since the spirit of Pad\'e approximation is to use a
ratio of polynomials and the power corrections $1/q^{2n}$ are
themselves in the canonical form of Pad\'e approximation. Let us
consider the following two-point function
\begin{eqnarray}
\Sigma(q^2) &=& q^2 \left( \log \frac{q^2}{\mu^2} + \frac{O_{2n}}{q^{2n}} \right) \nonumber \\
          &=& \Sigma_0(q^2) + \Sigma_{2n}(q^2).
\end{eqnarray}
Since we already know that the Pad\'e approximant to $\Sigma_0$ is
$q^2 R_\infty/Q_\infty$, as discussed in the previous section, and the correction $\Sigma_{2n}$
itself is already in the form of a Pad\'e approximant, we conclude that
the Pad\'e approximant $q^2 R(q^2)/Q(q^2)$ to $\Sigma(q^2)$ is
\begin{equation}
Q(q^2) = q^{2n} Q_\infty(q^2), \quad R(q^2)= q^{2n} R_\infty(q^2) +O_{2n}\,Q_\infty(q^2).
\end{equation}
Given the expression for $Q_\infty$ and $R_\infty$ in (\ref{qn3}) and (\ref{pn}) and
the identifications in (\ref{pade1}),
we obtain for $\Sigma(q^2)$
\begin{equation}
\Sigma(q^2) = q^2 \frac{q^{2n} J_0(qz_0) \log (q^2\epsilon^2) -
             \pi \ q^{2n}\, Y_0(qz_0) +  O_{2n} J_0(qz_0)}
        { q^{2n} J_0(qz_0)}.
\end{equation}
Therefore we see that the Pad\'e approximation reproduces the result
from AdS/CFT in (\ref{2pt4}) in the limit $q z_0 \gg 1$, identifying 
$O_{2n}= -\alpha O_{2n}^\prime$.  
In our approach, the lowest order in $O_{2n}$ simply yields a pole 
of order $n$ at $q^2=0$ to the Pad\'e approximant which obviously 
does not change the high energy behavior.
However, the perturbative approach 
that we used here is not valid for $q z_0 \approx 1$, and thus the
expression above including the spurious $q^2=0$ poles is not a valid 
description of the low-lying mass spectrum.
Indeed, in \cite{Hirn:2005vk} the shift in the mass for the low-lying
hadrons are computed numerically and found to be 
non-zero.\footnote{We thank Johanness Hirn
and Veronica Sanz for pointing out an erroneous statement about
the decay constants in an early version of this paper.}

\section{Physical Interpretation}
\label{sec:deconstruct}

We have shown that Migdal's approach using the Pad\'e
approximation reproduces the two-point function obtained from
Maldacena's AdS/CFT correspondence. In both approaches the goal
was to calculate at strong coupling; what is intriguing is that
they both lead to the same approximation of the vector
current-current correlator. In this section, we propose an
explicit correspondence between these two approaches.

We propose that finite $N$ in the Pad\'e approximation is
equivalent to $N$ hidden local symmetries
\cite{Bando:1984ej,Bando:1987br} that mock up the two-point
function resonances, which in turn is just an $N$-site
deconstruction \cite{Arkani-Hamed:2001ca,Hill:2000mu,
Sfetsos:2001qb,Falkowski:2002cm,Randall:2002qr} of the AdS space.
In the context of QCD, the idea of considering a large number of
hidden local symmetries was discussed in \cite{Son:2003et} and
found to qualitatively and in some cases quantitatively agree with
low energy data. Here our observation is that the poles in the
Pad\'e approximation can be interpreted as vector resonances. A
Pad\'e approximant at finite $N$ order has $N$ poles, which
correspond to $N$ vector resonances. Using hidden local symmetry,
each massive vector is reinterpreted as a massive gauge boson of a
broken gauge group. Hence, $N$ massive vector resonances can be
represented as a product gauge theory with $N$ broken gauge
groups.  Furthermore, the residues at the poles of the resonances
are positive \cite{Migdal:1977nu}, and can be understood as the
decay constants of the resonances.

According to our proposal, the resonance masses as determined by
the Pad\'e approximation become the Kaluza-Klein masses of a gauge
field in an extra dimension, so the product gauge theory described
above is equivalent to a deconstructed extra dimension in the
limit of a large number of lattice sites.  In the deconstructed
theory, the inverse lattice spacing roughly corresponds to the the
UV scale $\mu$, while the length of the AdS space corresponds to
$N/\mu$. The limits (\ref{limits-eq}) are equivalent to holding
the length of the AdS space fixed while taking the lattice spacing
to zero, which in deconstruction corresponds to obtaining a
continuous extra dimension.
The masses as determined by the Pad\'e approximation are given by
roots of Jacobi polynomials, which may differ from a
straightforward latticization of a slice of AdS$_5$. Hence, the
regularization of the 5D gauge theory by a moose model based on
the Pad\'e approximation differs from those discussed in
\cite{Sfetsos:2001qb,Abe:2002rj,Falkowski:2002cm,Randall:2002qr}, 
although
only in the way in which it approaches the continuum theory.
Nevertheless, these identifications, as well as the fact that the
spectrum obtained from the Pad\'e approximations matches the
Kaluza-Klein spectrum of a 5D gauge field in a slice of the
AdS$_5$ geometry, suggest it is natural to identify the large $N$
and low energy limit of the Pad\'e approximation with the
continuum limit of a deconstructed radial direction of AdS$_5$.

In AdS/QCD, a puzzle one might raise is the assumption that QCD,
being asymptotically free and renormalizable, is equivalent to a
5D gauge theory, which is neither asymptotically free nor
renormalizable. In this regard, Pad\'e approximation in the large
$N$ and low energy limit provides a UV completion of the 5D gauge
theory, since the linear moose theory is both asymptotically free
and renormalizable. In other words, Pad\'e approximation provides
a gauge invariant regularization of the 5D theory in which the
masses are given by roots of Jacobi polynomials.

We should point out that the argument for the emergence of a
deconstructed extra dimension is stronger than a simple
identification of vector boson masses.  The AdS/CFT prescription
for computing the generating functionals in the CFT arises from
the deconstructed extra dimension in a natural way, as was first
pointed out in \cite{Son:2003et}.  As a result, the deconstructed
theory that emerges from the Pad\'e approximation also predicts
identical decay constants as the AdS/CFT. Consider the following
action for the $N$-site linear moose model
\begin{equation}
\label{moose} S_{\rm m}= \int d^4x \sum_{n=1}^{N} v_n^2 \left|
D_\mu \Sigma^n \right|^2 - \sum_{n=1}^N
        \frac1{4 g_n^2} \left( F_{\mu\nu}^n \right)^2,
\end{equation}
where the covariant derivative is defined as
\begin{equation}
D_\mu \Sigma^n = \partial_\mu \Sigma^n - i A_\mu^{n-1} \Sigma^n +
i \Sigma^n A_\mu^n
\end{equation}
with $A_\mu^0=A_\mu^{N+1}=0$.
The current-current correlators can be computed by first gauging
the global symmetry on the zero$^{\rm th}$ site, where the CFT
lives,
and then differentiating the action 
with respect to the gauge field $A_\mu^0 \equiv B_\mu$:
\begin{equation}
\label{func}
\langle J_\mu(x) J_\nu(y) \rangle
   \equiv i \langle 0 | T( J_\mu(x) J_\nu(y) ) |0 \rangle
     = - \frac{\delta^2}{\delta B_\mu(x) \delta B_\nu(y)} \int {\cal D}A\ e^{i S_{\rm m}[A,B]}
\end{equation}
where the gauge field at the boundary $A_\mu^0=B_\mu$ is non-dynamical. In the continuum
limit this is equivalent to having a 5D gauge field $A_\mu(x,z)$ with
a fixed boundary condition
\begin{equation}
\label{bound}
A_\mu(x,0) = B_\mu(x).
\end{equation}
Furthermore, the action of the $N$-site moose $S_{\rm m}$ becomes
that of a 5D gauge theory in a slice of AdS$_5$. At tree-level,
the generating functional in (\ref{func}) is simply
\begin{equation}
\int {\cal D}A\ e^{i S_{\rm m}[A]} \approx e^{iS_{\rm
cl}[A_\mu^{\rm cl}]}
\end{equation}
where $A_\mu^{\rm cl}$ is the solution to the classical field
equation with the specified boundary condition (\ref{bound}). This
is the same as the AdS/CFT prescription in (\ref{pres}): the
generating functional for an operator in the CFT is the same as
the classical action of a bulk field whose boundary value serves
as the source for the operator.

The use of hidden local symmetry also makes it clear why a global
symmetry carried by the vector resonances on the CFT side must
become a gauge symmetry in the AdS picture. It is because each of
these vector resonances realizes one copy of the hidden local
symmetry at each lattice site. Therefore in the continuum limit
the gauge symmetry at each site becomes the gauge symmetry
 in the continuous extra dimension,
which is the bulk of AdS. It is worth emphasizing that the equivalence of these
two viewpoints is not a consequence of conformal symmetry, as exemplified by the discussion
in the previous section when we break the conformal symmetry by adding power corrections.

\section{Conclusions and Discussion}
\label{sec:conclusions}

Following suggestions by Shifman \cite{Shifman:2005zn} and Voloshin
 \cite{voloshin}, we have
studied the striking similarity between two approaches to estimating
the bound state spectrum and decay constants in asymptotically
conformal field theories: one that was invented by Migdal in the '70s and
the other making use of the AdS/CFT correspondence.
The latter approach was recently applied to QCD, the input being the
AdS/CFT correspondence and
chiral symmetry breaking.  The similarity of results based on holography
and results based on
other approximations may provide additional insight into the
predictive success of the AdS/QCD models.

In Migdal's Pad\'e approximation to the conformal two-point
correlation functions, the $N^{\rm th}$ order approximation gives
rise to a series of $N$ poles whose locations are
interpreted as resonance masses; the
residues of those poles are related to the decay constants of
those resonances.   Migdal found that the approach to
the logarithmic branch cut as $N$ is increased is through poles whose locations
are proportional to zeroes of Jacobi
polynomials, which approach Bessel functions in the large $N$
limit. Power corrections to the correlation function at leading order
do not modify the spectrum or the decay constants in the limit
$qz_0 \gg 1$. In the AdS/CFT approach, the geometry is
approximated by a slice of 5D anti-de Sitter space, whose isometry
near the AdS boundary reproduces the conformal symmetry of the
corresponding field theory in the UV.  Solutions to the equations
of motion for 5D fields correspond to resonances. The AdS
geometry leads to vector meson masses that are zeroes of Bessel functions,
which match the result of the Pad\'e approximation to the
conformal two-point functions after identifying the IR
scales in both approaches. Power corrections are included by
modification of the AdS geometry, which leads to a modification of
the decay constants once again in agreement with the Pad\'e
approximation.

Migdal's approach is similar in spirit to QCD sum rules
\cite{Shifman:1978bx,Shifman:1978by}: the input is high energy
information about QCD in the form of the operator product
expansion, together with some assumptions about the dynamics of
chiral symmetry breaking and confinement; and the output is low
energy information about hadrons such as their spectrum, decay
constants and couplings. AdS/QCD is similar in spirit as well,
where the prescription for extrapolating between UV and IR is
dictated by the dynamics of a higher dimensional field theory in
AdS space. We find it remarkable that Migdal's method via the
Pad\'e approximation is exactly equivalent to the holographic
prescription from AdS/CFT, at least for the vector current
two-point function.

We mainly concentrated on two-point functions of vector currents
in this paper. It is natural to ask whether the equivalence
between the approaches of Migdal and AdS/CFT persists beyond
two-point functions\footnote{See \cite{Dosch:1977qh}, for example,
for an attempt to generalize Migdal's approach to three-point
amplitudes.} and if from Migdal's approach one could derive the
higher dimensional theory in a systematic fashion. In terms of
extending the range of validity of a holographic dual of QCD in a
phenomenological approach such as AdS/QCD, there are still many
hurdles to overcome. For example, a simple argument (see for
example \cite{Shifman:2005zn}) for the Regge trajectory of QCD
suggests that at high energies the mass of the $N^{\rm th}$
excited hadron is proportional to $\sqrt{N}$, whereas in AdS/QCD
models the mass is generically proportional to $N$ for highly
excited states. On the other hand, the bulk metric must always
approach AdS near the boundary in order to reproduce the conformal
symmetry. It would be interesting to find an AdS/QCD model that
would give rise to the required Regge trajectory and the
asymptotic conformal behavior at the same time.

In Sec. \ref{sec:deconstruct} we saw that, by combining the ideas
of hidden local symmetry and deconstruction, the resonances
determined by the Pad\'e approximation can be interpreted as
arising from a linear moose model.  We discussed that the moose
model becomes a deconstructed warped extra dimension when the
number of resonances becomes large. We further showed that this
extra dimension can be identified with the radial direction of
anti-de Sitter space. It is interesting to ask what would happen
if we applied the same setup to the two-point function of
stress-energy tensors in the CFT. In this case, there is a tower
of massive spin-2 resonances in the Pad\'e approximation. A
massive spin-2 particle can be thought of as a massive graviton,
realizing a copy of broken 4D diffeomorphism invariance. Therefore
a theory for $N$ massive spin-2 resonances might be interpreted in
terms of an $N$-site latticized extra dimension with gravity.
However, a deconstructed gravitational theory is very different
from a deconstructed gauge theory as the latticized gravity
suffers from a strong coupling problem
 \cite{Arkani-Hamed:2002sp} in such way that the effective
$N$-site theory breaks down at a much lower scale than might
naively be expected. It was recently found that in the AdS space,
the strong coupling problem is not as severe as in flat space
\cite{Randall:2005me,Gallicchio:2005mh}. It would be interesting
to consider strong coupling issues from the perspective of the
Pad\'e approximation. An approach like this for the spin-2
resonances might lead to a better understanding of how a
gravitational theory can be equivalent to a field theory in fewer
dimensions without gravity.

\section*{Acknowledgments}
This work was inspired by Shifman and Voloshin's comments on
Migdal's work, which we acknowledge. We benefited from
conversations with Juan Maldacena, Veronica Sanz, and Lior
Silberman. We are also grateful to Chris Carone for collaboration at early
stages of this work. 
Correspondence with Johanness Hirn and Veronica Sanz on Sec. 4
is acknowledged.  
We thank Martin Schmaltz for discussions leading to clarification
of our derivation of Eq.~(\ref{qn2}). 
This work is supported in part by the National Science
Foundation under grant PHY-0504442 and the Jeffress Memorial Trust
under grant J-768 (JE), and by the Department of Energy under
grants DE-FG02-96ER40969 (GDK) and DE-FG02-90ER40542 (IL). Part of
this work was performed at the Aspen Center for Physics, as well
as during various visits to IAS and the College of William and Mary,
whose hospitalities we acknowledge.

\section*{Appendix}

In this appendix we provide a derivation of (\ref{qn3}). We make
use of the identity \cite{szego}
\begin{equation}
P_n^{(a,b)}(x) = \frac{\Gamma(n+a+1)}{\Gamma(n+1)\Gamma(a+1)}
    \left(\frac{1+x}2 \right)^n\phantom{1}_2F_1\left(-n,-n-b\,;a+1;\frac{x-1}{x+1}\right),
\end{equation}
where the hypergeometric function is defined as
\begin{equation}
_2F_1(a,b;c;x) = \sum_{m=0}^{\infty} \frac{\Gamma(a+m)}{\Gamma(a)}
                \frac{\Gamma(b+m)}{\Gamma(b)} \frac{\Gamma(c)}{\Gamma(c+m)}
               \  \frac{x^m}{m!}.
\end{equation}
In (\ref{qn2}) we need to consider
\begin{eqnarray}
\label{qn4}
 (1+s)^N P_N^{(a,b)}\left(\frac{1-s}{1+s}\right) &=&   \frac{\Gamma(N+a+1)}{\Gamma(N+1)\Gamma(a+1)}
                  \phantom{1}_2F_1(-N,-N-b\,;a+1;-s) \nonumber \\
              &=& \sum_{m=0}^{\infty} \frac{(-s)^m}{\Gamma(a+1+m) m!}\
                \frac{\Gamma(N+a+1)\Gamma(N+b+1)}{\Gamma(N-m+1)\Gamma(N+b+1-m)}
\end{eqnarray}
where we have used the identity
\begin{equation}
\frac{\Gamma(-N+m)}{\Gamma(-N)} = (-1)^m \frac{\Gamma(N+1)}{\Gamma(N-m+1)}.
\end{equation}
In the $N\to \infty$ limit we can make use of the Sterling's formula
\begin{equation}
\lim_{x\to\infty} \Gamma(x) = e^{-x} x^{x-1/2} \sqrt{2\pi}.
\end{equation}
Then it is straightforward to show that in (\ref{qn4})
\begin{equation}
 \lim_{N\to\infty} \frac{\Gamma(N+a+1)\Gamma(N+b+1)}{\Gamma(N-m+1)\Gamma(N+b+1-m)}
 \longrightarrow N^{2m+a} \quad ,
 \end{equation}
which leads to
\begin{eqnarray}
Q_a(t) &=& N^a \sum_{m=0}^\infty \frac{(-4N^2t)^m}{2^{2m} \Gamma(a+m+1) m!}
   \nonumber \\
   &=&  t^{-a/2} J_a(2N\sqrt{t}).
\end{eqnarray}
For $a=0$ we obtain (\ref{qn3}).



\begin{thebibliography}{99}

\baselineskip=-10mm

\bibitem{Maldacena:1997re}
  J.~M.~Maldacena,
  Adv.\ Theor.\ Math.\ Phys.\  {\bf 2}, 231 (1998)
  [arXiv:hep-th/9711200].

\bibitem{Gubser:1998bc}
  S.~S.~Gubser, I.~R.~Klebanov and A.~M.~Polyakov,
  Phys.\ Lett.\ B {\bf 428}, 105 (1998)
  [arXiv:hep-th/9802109].

\bibitem{Witten:1998qj}
  E.~Witten,
  Adv.\ Theor.\ Math.\ Phys.\  {\bf 2}, 253 (1998)
  [arXiv:hep-th/9802150].

\bibitem{Arkani-Hamed:2000ds}
  N.~Arkani-Hamed, M.~Porrati and L.~Randall,
  JHEP {\bf 0108}, 017 (2001)
  [arXiv:hep-th/0012148].

\bibitem{Randall:1999ee}
  L.~Randall and R.~Sundrum,
  Phys.\ Rev.\ Lett.\  {\bf 83}, 3370 (1999)
  [arXiv:hep-ph/9905221].

\bibitem{Randall:1999vf}
  L.~Randall and R.~Sundrum,
  Phys.\ Rev.\ Lett.\  {\bf 83}, 4690 (1999)
  [arXiv:hep-th/9906064].

\bibitem{Sakai:2004cn}
  T.~Sakai and S.~Sugimoto,
  Prog.\ Theor.\ Phys.\  {\bf 113}, 843 (2005)
  [arXiv:hep-th/0412141].

\bibitem{deTeramond:2005su}
  G.~F.~de Teramond and S.~J.~Brodsky,
  Phys.\ Rev.\ Lett.\  {\bf 94}, 201601 (2005)
  [arXiv:hep-th/0501022].

\bibitem{Erlich:2005qh}
  J.~Erlich, E.~Katz, D.~T.~Son and M.~A.~Stephanov,
  arXiv:hep-ph/0501128.

\bibitem{DaRold:2005zs}
  L.~Da Rold and A.~Pomarol,
  Nucl.\ Phys.\ B {\bf 721}, 79 (2005)
  [arXiv:hep-ph/0501218].

\bibitem{Hirn:2005nr}
  J.~Hirn and V.~Sanz,
  JHEP {\bf 0512}, 030 (2005)
  [arXiv:hep-ph/0507049].

\bibitem{Sakai:2005yt}
  T.~Sakai and S.~Sugimoto,
  Prog.\ Theor.\ Phys.\  {\bf 114}, 1083 (2006)
  [arXiv:hep-th/0507073].

\bibitem{DaRold:2005vr}
  L.~Da Rold and A.~Pomarol,
  arXiv:hep-ph/0510268.

\bibitem{Ghoroku:2005vt}
  K.~Ghoroku, N.~Maru, M.~Tachibana and M.~Yahiro,
  Phys.\ Lett.\ B {\bf 633}, 602 (2006)
  [arXiv:hep-ph/0510334].



\bibitem{Hirn:2005vk}
  J.~Hirn, N.~Rius and V.~Sanz,
  arXiv:hep-ph/0512240.

\bibitem{Son:2003et}
  D.~T.~Son and M.~A.~Stephanov,
  Phys.\ Rev.\ D {\bf 69}, 065020 (2004)
  [arXiv:hep-ph/0304182].

\bibitem{Brodsky:2003px}
  S.~J.~Brodsky and G.~F.~de Teramond,
  Phys.\ Lett.\ B {\bf 582}, 211 (2004)
  [arXiv:hep-th/0310227].

\bibitem{Evans:2004ia}
  N.~J.~Evans and J.~P.~Shock,
  Phys.\ Rev.\ D {\bf 70}, 046002 (2004)
  [arXiv:hep-th/0403279].

\bibitem{Hong:2004sa}
  S.~Hong, S.~Yoon and M.~J.~Strassler,
  arXiv:hep-th/0409118.

\bibitem{Shifman:2005zn}
  M.~Shifman,
  arXiv:hep-ph/0507246.

\bibitem{voloshin}
M.~B.~Voloshin, private communication.

\bibitem{Migdal:1977nu}
A.~A.~Migdal,
 Annals Phys.\  {\bf 109}, 365 (1977).


\bibitem{Migdal:1977ut}
  A.~A.~Migdal,
  Annals Phys.\  {\bf 110}, 46 (1978).

\bibitem{Leigh:2005dg}
  R.~G.~Leigh, D.~Minic and A.~Yelnikov,
  arXiv:hep-th/0512111.

\bibitem{Shifman:1978bx}
  M.~A.~Shifman, A.~I.~Vainshtein and V.~I.~Zakharov,
  Nucl.\ Phys.\ B {\bf 147}, 385 (1979).


\bibitem{Shifman:1978by}
  M.~A.~Shifman, A.~I.~Vainshtein and V.~I.~Zakharov,
  Nucl.\ Phys.\ B {\bf 147}, 448 (1979).



\bibitem{Arkani-Hamed:2001ca}
N.~Arkani-Hamed, A.~G.~Cohen and H.~Georgi,
Phys.\ Rev.\ Lett.\  {\bf 86}, 4757 (2001) [arXiv:hep-th/0104005].

\bibitem{Hill:2000mu}
C.~T.~Hill, S.~Pokorski and J.~Wang,
Phys.\ Rev.\ D {\bf 64}, 105005 (2001) [arXiv:hep-th/0104035].



\bibitem{szego}
See, for example,  G.~Szego, ``Orthogonal Polynomials,'' American Mathematical
Society Colloquium Publications, Volume XXIII, 1975.

\bibitem{Balasubramanian:1998sn}
  V.~Balasubramanian, P.~Kraus and A.~E.~Lawrence,
  Phys.\ Rev.\ D {\bf 59}, 046003 (1999)
  [arXiv:hep-th/9805171].


\bibitem{Balasubramanian:1998de}
  V.~Balasubramanian, P.~Kraus, A.~E.~Lawrence and S.~P.~Trivedi,
  Phys.\ Rev.\ D {\bf 59}, 104021 (1999)
  [arXiv:hep-th/9808017].



\bibitem{Randall:2001gb}
  L.~Randall and M.~D.~Schwartz,
  JHEP {\bf 0111}, 003 (2001)
  [arXiv:hep-th/0108114].

  \bibitem{sanz}
  V. Sanz, private communication.

\bibitem{Bando:1984ej}
M.~Bando, T.~Kugo, S.~Uehara, K.~Yamawaki and T.~Yanagida,
Phys.\ Rev.\ Lett.\  {\bf 54}, 1215 (1985).

\bibitem{Bando:1987br}
  M.~Bando, T.~Kugo and K.~Yamawaki,
  Phys.\ Rept.\  {\bf 164}, 217 (1988).

\bibitem{Sfetsos:2001qb}
  K.~Sfetsos,
  Nucl.\ Phys.\ B {\bf 612}, 191 (2001)
  [arXiv:hep-th/0106126].

\bibitem{Abe:2002rj}
 H.~Abe, T.~Kobayashi, N.~Maru and K.~Yoshioka,
 Phys.\ Rev.\ D {\bf 67}, 045019 (2003)
 [arXiv:hep-ph/0205344].



\bibitem{Falkowski:2002cm}
  A.~Falkowski and H.~D.~Kim,
  JHEP {\bf 0208}, 052 (2002)
  [arXiv:hep-ph/0208058].

\bibitem{Randall:2002qr}
  L.~Randall, Y.~Shadmi and N.~Weiner,
  JHEP {\bf 0301}, 055 (2003)
  [arXiv:hep-th/0208120].

\bibitem{Dosch:1977qh}
  H.~G.~Dosch, J.~Kripfganz and M.~G.~Schmidt,
  Phys.\ Lett.\ B {\bf 70}, 337 (1977).

\bibitem{Arkani-Hamed:2002sp}
  N.~Arkani-Hamed, H.~Georgi and M.~D.~Schwartz,
  Annals Phys.\  {\bf 305}, 96 (2003)
  [arXiv:hep-th/0210184].

\bibitem{Randall:2005me}
  L.~Randall, M.~D.~Schwartz and S.~Thambyapillai,
  JHEP {\bf 0510}, 110 (2005)
  [arXiv:hep-th/0507102].

\bibitem{Gallicchio:2005mh}
  J.~Gallicchio and I.~Yavin,
  arXiv:hep-th/0507105.


\end{thebibliography}
\end{document}